# Self-organisation of highly symmetric nanoassemblies: a matter of competition

Jesus E. Galván-Moya[†,#], Thomas Altantzis[‡,#], Kwinten Nelissen[†], Francois M. Peeters[†], Marek Grzelczak[§,‖], Luis M. Liz-Marzán[§,‖], Sara Bals[‡]*, Gustaaf Van Tendeloo[‡]

[†]CMT, University of Antwerp, Groenenborgerlaan 171, B-2020 Antwerp, Belgium

[‡]EMAT, University of Antwerp, Groenenborgerlaan 171, B-2020 Antwerp, Belgium

[§]Bionanoplasmonics Laboratory, CIC biomaGUNE, Paseo de Miramón 182, 20009 Donostia-San Sebastián, Spain

[‖]Ikerbasque, Basque Foundation for Science, 48011 Bilbao, Spain

## AUTHOR INFORMATION

### # Author contribution

J.E.G-M and T.A. equally contributed to this work.

### * Corresponding author

Email: sara.bals@uantwerpen.be

The authors declare no competing financial interest

## ASSOCIATED CONTENT

### Supporting information

Details concerning synthesis and structural analysis, extended theoretical analysis and material dependence explanation. This material is available free of charge via the internet at http:// pubs.acs.org.

**ABSTRACT**

The properties and applications of metallic nanoparticles are inseparably connected not only to their detailed morphology and composition, but also to their structural configuration and mutual interactions. As a result, the assemblies often have superior properties as compared to individual nanoparticles. Although it has been reported that nanoparticles can form highly symmetric clusters, if the configuration can be predicted as a function of the synthesis parameters, more targeted and accurate synthesis will be possible. We present here a theoretical model that accurately predicts the structure and configuration of self-assembled gold nanoclusters. The validity of the model is verified using quantitative experimental data extracted from electron tomography 3D reconstructions of different assemblies. The present theoretical model is generic and can in principle be used for different types of nanoparticles, providing a very wide window of potential applications.

Assemblies of nanoparticles in two and three dimensions have gained increasing interest because of their multiple applications[1-5] and improved properties, compared to those of their building blocks.[6-10] By varying experimental parameters, such as the size and shape of the individual particles or the nature and length of the capping ligands, different nanostructures with unique configurations can be obtained. However, the complex mechanism and interplay of the forces and processes leading to a given assembly is often poorly understood and in some cases only empirically found. Having access to such information would enable researchers to predict the stacking of the individual nanoparticles into a specific configuration as a function of the experimental parameters. In this manner, the synthesis of nanoassemblies with tailored properties for specific applications would become much more accurate and efficient. In this letter, we demonstrate a thorough understanding of the formation of symmetrical 3D nanoassemblies by combining a state-of-the-art structural analysis with modern computational techniques.

Transmission Electron Microscopy (TEM) is an ideal technique to investigate materials at the (sub)nanometer scale and has therefore been widely used in the study of nanomaterials. In order to understand the connection between structure and properties of these materials, the combination of TEM and theoretical calculations is very powerful.[11-13] However, it is important to realize that TEM images only correspond to a two-dimensional (2D) projection of a three-dimensional (3D) object. In order to gain the necessary structural information concerning the 3D nanoassemblies, 3D TEM, so-called electron tomography, has to be performed.[14-16] Recently, this technique has proven its power in the investigation of 3D nanoassemblies, especially when quantitative data, such as particle diameters or positions, are required.[1,16] 2D self-assembled systems have been theoretically studied in depth during the last decade,[17-23] and the transition between 2D and 3D assemblies has been recently investigated.[22] However, these studies are mainly based on phenomenological models,[1,11,25-28] but do not yield insight concerning the underlying physical processes involved during the formation of the 3D assemblies.

In the present work, we combined state-of-the-art electron tomography results with a new theoretical model, which is easy to implement and moreover, which provides a thorough understanding of the formation of 3D assemblies from the aggregation of gold nanoparticle building blocks. In the selected example shown in Figure 1, gold nanospheres capped with polystyrene chains form clusters in solution upon increasing the solvent dielectric constant. This was interpreted as an interplay between van der Waals and hydrophobic attraction and steric repulsion forces. The model we propose here is more general, in the sense that it is based on the competition between a long-range attractive and a short-range repulsive force for each pair of particles. This is explained as an effective interaction between particles, which in the present study will be taken as isotropic, assuming that all nanoparticles are identical and spherical, which is reasonable in the case of the experiments considered here. The combination of electron tomography and the new theoretical model enables us to predict the 3D configuration of the obtained Au nanoassemblies with high accuracy. It is however important to point out that, because of the generality of the model, our approach can be applied to a wide variety of nanoparticle assemblies.

Using electron tomography, the 3D configurations were determined for several assemblies obtained using different synthesis parameters. The most relevant parameters correspond to the diameter of the individual nanoparticles (D) and the polymer chain length (L). The experimental results are presented in Figure 1; they show essentially two different kinds of configurations. For assemblies containing particles with a relatively small diameter capped with polymers of short chain length (L) a dense packed configuration was found (Figure 1 a-c). However, in the case of longer polymer lengths, shell like structures can be identified (Figure 1 d-i). Strikingly, some of the 3D reconstructions, such as the example presented in Figure 1d yield a highly symmetric and regular 3D stacking of the individual nanoparticles. In Figure 1g, an icosahedron is clearly observed, but also other types of polyhedra were found. It should be noted that for large and more dense packed assemblies, such as the example in Figure 1f, regular stackings were not observed, but the arrangement was closer to spherical symmetry. For certain configurations however, some shells appear to be incomplete, i.e. particles are missing. In order to obtain a thorough understanding of the formation of such assemblies, the electron tomography results were compared with computer simulations.

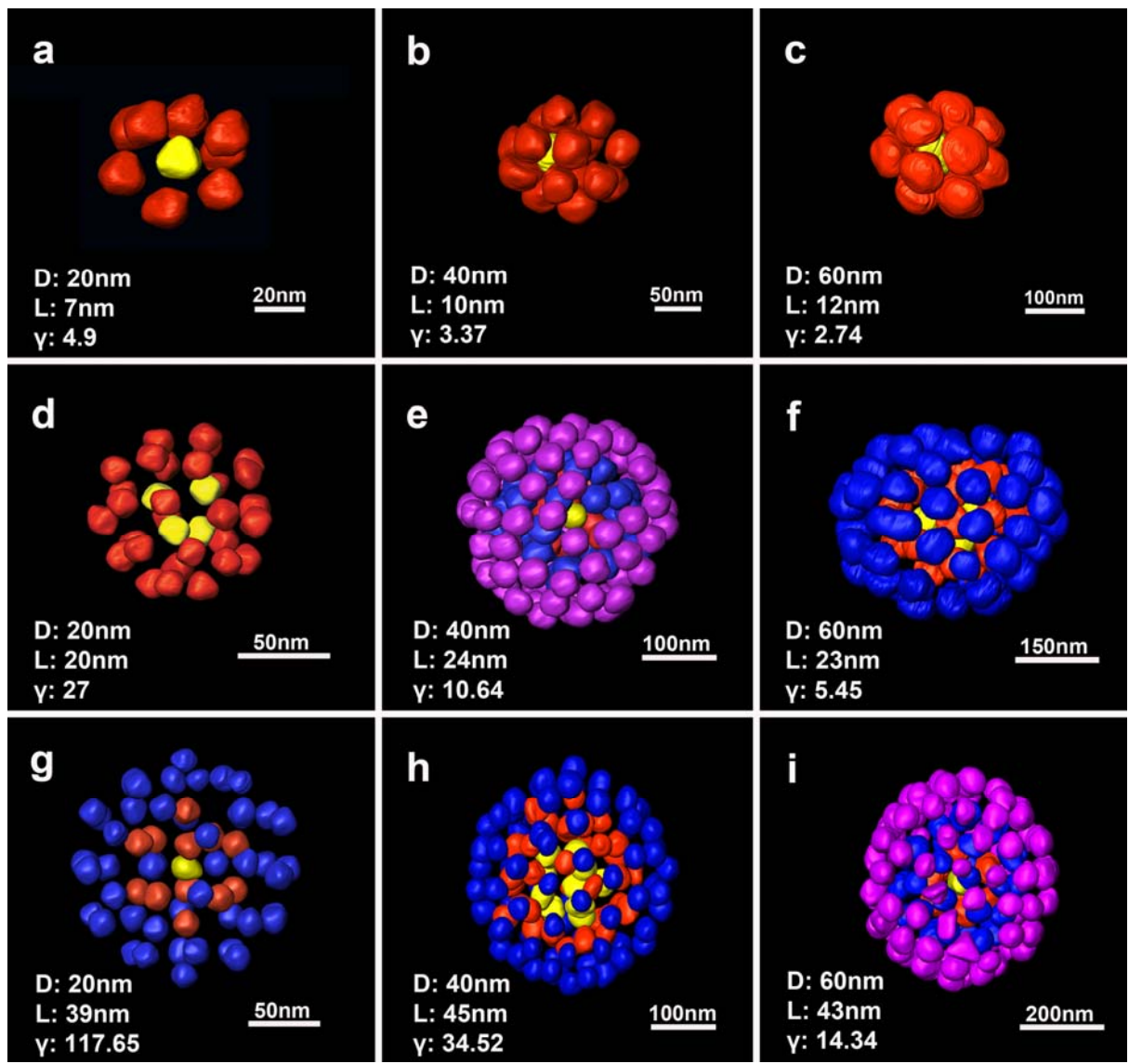

**Figure 1.** 3D representation of reconstructed assemblies from experimental 2D TEM images at various angles. In each image, the synthesis parameters (D and L) and the value of $\gamma$ for each configuration are given. Different colors refer to the different shells in the assemblies.

The inter-particle distances in such self-organized systems have been explained theoretically, showing that the interaction between particles is complex and involves many variables.[1,28] Although the model proposed in Reference 1 may very well predict inter-particle distances, it turns out that the 3D configuration of the assemblies can not be obtained. The reason is that in the model of Reference 1, the van der Waals interaction is only correctly described in the limit of close-approach and overestimates the strength of the long-range interaction. Here, we propose a potential that is based on a generalized Morse inter-particle interaction:

$$E = \sum_{i=1}^{N} \sum_{j>i}^{N} \left( A \exp(-\alpha\, r_{ij}) - \widetilde{B} \exp(-\widetilde{\beta}\, r_{ij}) \right). \tag{1}$$

In this expression A ($\widetilde{B}$) and $\alpha$ ($\widetilde{\beta}$) are real numbers that modulate the strength and the screening of the repulsion (attraction) between particles, $N$ is the number of particles in the assembly, and $r_{ij}$ represents the distance between the centers of the $i$-th and the $j$-th particles in the self-assembled system. A similar interaction was used previously to describe the structure of 2D self-organized colloidal systems.[29] For a given temperature T, Equation (1) can be written in dimensionless form as follows:

$$E = \sum_{i=1}^{N} \sum_{j>i}^{N} \left( \exp(-r_{ij}) - B \exp(-\beta\, r_{ij}) \right), \tag{2}$$

where $\beta = \widetilde{\beta}/\alpha$, while the energy and the distances are given in units of $E_0 = A K_B T$ and $r_0 = \alpha \widetilde{r}_0$, respectively. The average distance between concentric shells in the assembly is given by $\widetilde{r}_0$, which is defined as the characteristic length of the system (see Supporting Information for details). Equation (2) represents a two-parameter model which enables one to tune the strength and the range of the attraction in a flexible manner.

The advantage of this potential is that, in the inter-particle distance range near its minimum, it can be mapped on the potential of Reference 1, from which experimental synthesis conditions can be extracted. Once this relationship has been established and the theoretical parameters have been linked to the synthesis conditions, the procedure can be reversed and thus, starting from experimental synthesis conditions the final configuration can be predicted.

The ground state configuration is obtained by Monte Carlo (MC) simulations supplemented with the Newton optimization method. This approach has been successfully used in previous works.[17,30,31] As a first result, irrespective of the values of $B$ and $\beta$, we notice that the model is translational and rotational invariant. Due to the isotropic interparticle interaction, highly symmetric structures were found, in agreement with our electron tomography results. Based on numerical simulations, a good agreement with the experimental results has been found with $\beta = 0.5$, for all samples investigated. This implies a relative short-range attraction between the particles. The parameter B can be used as an adjustable parameter that determines the packing density. A detailed discussion on the relation between the present model and the synthesis parameters can be found in the Supporting Information.

In order to correlate the experimental data with the theoretical calculations, we introduce the mass density inside the shell defined by the ends of the polymer chains surrounding each gold nanoparticle, which is given by $\rho = \rho_c / \gamma$, where $\rho_c$ is the density of Au and

$$\gamma = \left(1 + \frac{2L}{D}\right)^3 \tag{3}$$

is the ratio of the total volume taken by the particle and the polymer and the volume of the Au particle, which is thus a material independent quantity. Experimentally, we found that for $\gamma \gg 1$, i.e. long polymer chain lengths with respect to the particle size, the assembly formed by the nanoparticles is highly spherical symmetric (see Figure 1). As can be seen from Equation (2), the parameter $B$ is a measure for the attraction between each pair of particles. As an example, we model three different clusters, which correspond to Figures 1 d, g, h. For the largest $\gamma$-values, as those listed in Table 1, the best fit with the tomographic reconstruction was found with $B = 0.65$.

| Sample | D(nm) | L(nm) | $\gamma$ | N | n | Shell-like Configuration | |
|---|---|---|---|---|---|---|---|
| | | | | | | Tomographic Reconstruction | Theoretical Prediction |
| A | 20 | 20 | 27.00 | 33 | 2 | (4, 29) | (4, 29) |
| B | 20 | 39 | 117.65 | 59 | 3 | (1, 12, 46) | (1, 12, 46) |
| C | 40 | 45 | 34.52 | 132 | 3 | (11, 37, 84) | (10, 37, 85) |

**Table 1.** Experimental sample parameters: D is the diameter of the Au particles and L is the length of the polymer chains surrounding them, N is the number of particles and $n$ is the number of shells found in each cluster. The number of particles per shell for the experimental assemblies and for the theoretical model with B=0.65 are displayed in the two rightmost columns.

The best agreement between theory and experiment was achieved for large $\gamma$-values, when complete shell structures, i.e. without vacancies, are formed experimentally. We therefore focus on the interpretation of the three different assemblies having large $\gamma$-values, with the experimental parameters listed in Table 1.

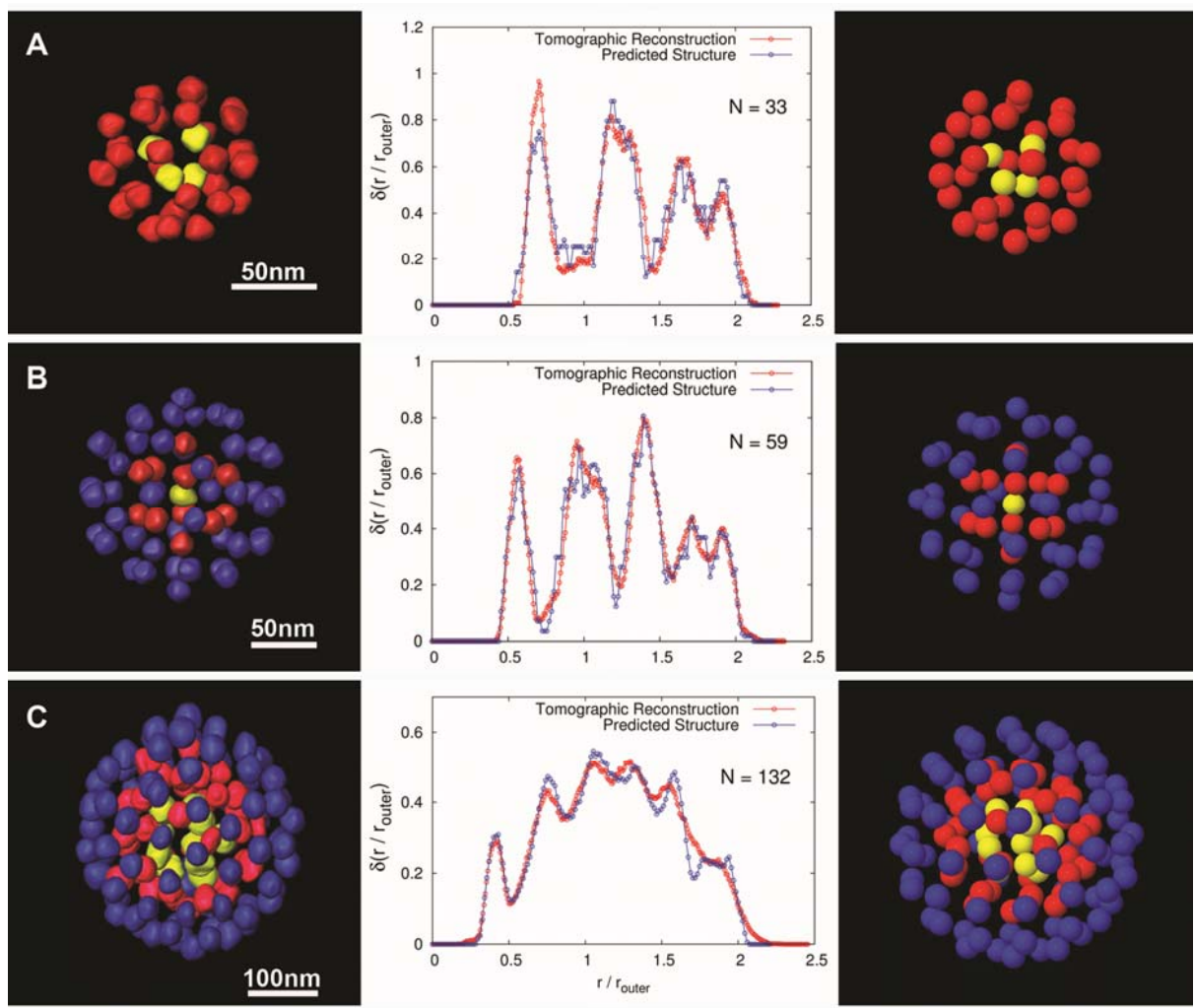

**Figure 2.** Comparison between configurations obtained from experimental reconstruction (left) and simulations (right) with the parameters listed in Table 1. Particles belonging to different shells are highlighted by different colors. In the central panel, the radial density distribution function is plotted, where we compare the experimental results with the ones obtained from theory.

Figure 2 demonstrates the excellent agreement between the theoretical predictions and the experimental configurations obtained by tomographic reconstruction. The comparison between experiment and simulations is based on the number of particles per shell (rightmost columns of Table 1). It can be seen that the simulations predict with high precision the particle positions for small and intermediate cluster sizes ($N = 33, 59$), whereas a small discrepancy of only one particle between the inner and the outer shell is found for the largest one ($N=132$). To further confirm the agreement we plotted the radial density distribution of the assemblies ($\delta(r)$) as a function of the interparticle distance, which for this figure is scaled by the radius of the outer shell ($r_{outer}$). This function is defined as the probability to find two particles separated by a distance $r$, and is closely related to the radial distribution function, which is used to describe the structure of the assemblies for larger systems.[13]

Based on the use of $\delta(r)$, a comparison is presented in the central panel of Figure 2. For the experimental data we used the coordinates of the center of mass of all nanoparticles in each assembly, as they were extracted from the tomographic reconstructions. The radial densities show very good agreement not only in the number of peaks, which is intrinsically related to the shell-like structure, but also with respect to the location and height of the peaks.

Also from the simulation results it is clear that the particles at the inner shells of the assemblies preferably sit in highly symmetric polyhedral structures. For example, the third column of Figure 2 confirms that the inner shell of sample A is formed by a regular tetrahedron, and the second shell of sample B forms a regular icosahedron (see Supporting Information). These are just few examples of regular structures that can be found. It must be noted that also octahedra and different elongated bipyramidal structures are predicted by our model, to form the inner shell configuration of assemblies with different numbers of particles. This finding is in contrast to Lennard-Jones assemblies, which were recently found to organize in planar configurations.[11]

Simulations carried out for assemblies with an increasing number of particles enabled us to describe the self-assembly process as follows: initially, for a small number of particles ($N<12$), all particles are arranged in a single shell, forming configurations such as antiprisms and bipyramids but with regular polyhedra (tetrahedron, octahedron and icosahedron) being dominant. Next, as more particles are added, the outer shell gets filled and the outer radius increases. Consequently, the inter-shell distances increase. After increasing the number of particles further it becomes energetically more favorable to occupy the next inwards located shell due to the high surface tension of the former shell. This process is repeated shell after shell till the most inner shell reaches a critical size leading to a void at the center of the assembly. After a further particle addition, this void is occupied by an additional particle, resulting in the formation of a new shell. Such assembly configurations can be grouped in Mendeleev-like tables as function of the number of particles, an example is given in the supporting information.

In order to illustrate the physics behind the 3D assembly process, we show in Figure 3 the phase diagram of the ground state configurations for a system with $N = 59$ particles as a function of the parameters in our model. In this figure, the different letters represent different configurations of the system, which are given at the right side of the figure by the number of particles in each shell. The thick solid arrow indicates the direction of increasing packing fraction. Please note that the best fit with the present experiment was obtained for $B = 0.65$ and $\beta = 0.5$. All transitions between different configurations were found to be of first order.

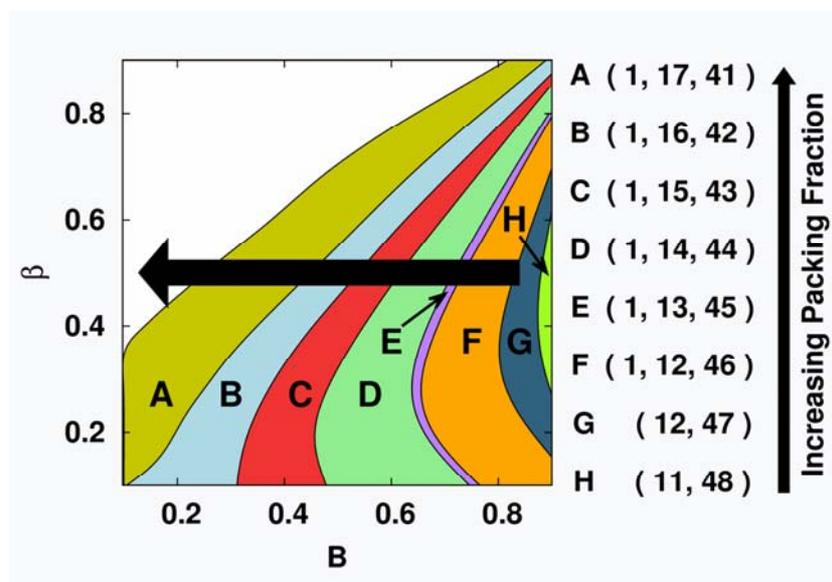

**Figure 3.** Phase diagram of the ground state configuration of an assembly with $N = 59$ gold nanoparticles, as a function of the theoretical parameters $B$ and $\beta$. Different colors represent different configurations as they are labeled alphabetically in the figure. The number of particles per shell for each configuration is indicated in the column on the right side. The thick solid arrow indicates the direction of increasing packing fraction.

Following the configurations as they are alphabetically ordered in Figure 3, one can observe that increasing the attraction ($B$) leads to the migration of particles from the inner to the outer shells. This is a consequence of the reduction of the inter-particle distance allowing the outer shells to accommodate more particles.

In Figure 3, the configurations at the left of A can no longer be described as shell-like structures; they are structures with planar faces, which is typical for particles interacting through a Lennard-Jones or Morse potential.[11,26] Starting from configuration A and by increasing parameter B, all configurations retain a shell-like structure after going through the following configurations: dense packed configuration → regular triangular configuration (polyhedral configurations) → spherical-like shell configuration. The last transition occurs through a continuous process of shell radius reduction.

Our theoretical approach allows us to explain the formation of the assemblies and to correctly predict their 3D configuration for the considered synthesis parameters of our samples. The observed structures result from the tendency of the system to form a close packed configuration, as obtained for small values of B, as well as from the formation of a shell-like structure due to strong attraction. The stronger the attraction, the more particles can be packed on a specific shell, and the more shell-like the final structure will be. This phenomenon is closely related to the surface tension, where the attraction between molecules or atoms results in the minimization of the surface formed by the particles on the outer shell. This competition results, in the case of strong attraction, in a sequential formation of regular polyhedra. Although shell-like structures are expected to form for large $\gamma$-values, experimental evidence showed that this is the case even in the region $27 \leq \gamma \leq 117$. Experimentally, close packed configurations are expected for nanoparticles with a weak attraction, while shell-like structures are expected to be formed for nanoparticles with a strong attraction.

In the formation process previously described, we found that the attraction between particles is of relative short-range character ($\beta = 0.5$). This allowed us to obtain an optimal agreement between theory and experiment (Figure 2), showing that the screening of the attraction is not affected by either the particle size or the polymer length. The relation between synthesis parameters and the model parameter B is obtained by fitting the potential presented in Reference 1 around the local minimum with our model (Equation (1)). In Figure 4 this relation is given for different values of particle size (D) and polymer length (L). All synthesis parameters are listed in the supporting information.

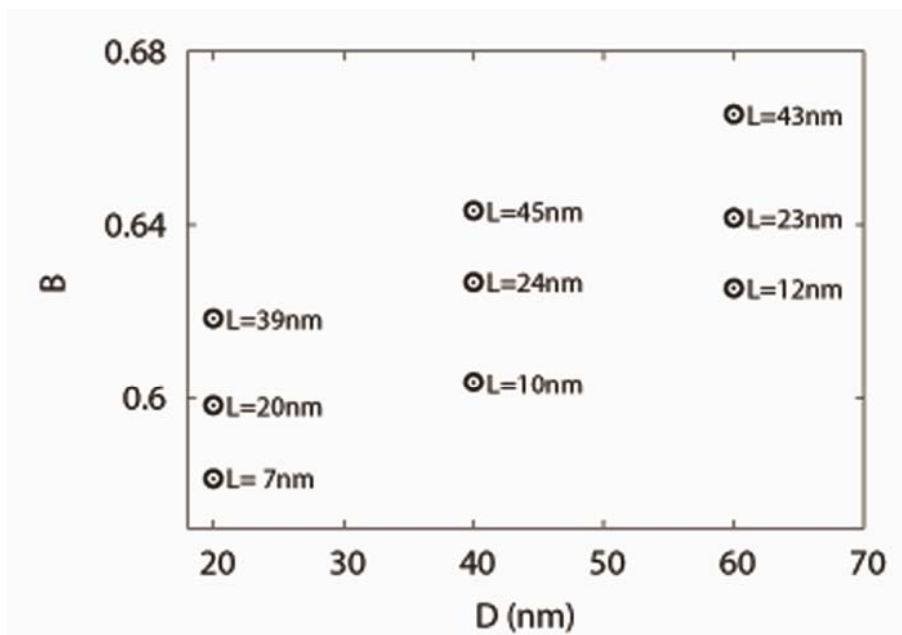

**Figure 4.** Theoretical parameter (B) plotted as a function of the particle diameter (D) for different values of the polymer chain length (L). These points have been calculated by adjusting the energy curve proposed in Reference 1 through Equation (1) for $\beta = 0.5$. A complete list of the synthesis parameters considered in the present calculation is shown in the supporting information.

From Figure 4 one can see that the value of B is proportional to both particle size and polymer chain length. From these results, one can conclude that the attraction between particles increases with the polymer length (which may be related to hydrophobic forces), but that the increment becomes less significant for smaller particle sizes (where van der Waals forces are weaker). Figure 4 reveals that the values of B are in the interval [0.58, 0.67] (see supporting information), which confirms the validity of the present model where the theoretical structures correspond to B=0.65 (see Figure 2). These results indicate that by increasing the polymer chain length the strength of the attraction between particles increases, resulting in a reduced packing fraction and a more symmetrical configuration.

Nanoparticles in the 10-100 nm range tend to self-organize into three dimensional clusters. In this letter, this self-organization process of spherical nanoparticles into spherical assemblies was investigated, both experimentally and theoretically, using a pairwise interaction; Au nanoparticle assemblies formed by inducing hydrophobic forces were chosen as a case study. We proposed a new model, based on a simple competition between attractive and repulsive interactions which we were able to relate with the experimental synthesis parameters. The excellent agreement between the experimentally observed and the theoretically predicted configurations provides us with the opportunity to understand the physics behind cluster formation. For the synthesis parameters used, we can explain the particular cluster formation and correctly predict their 3D configuration. The final structures result from the tendency of the system to form a close packed configuration and the formation of a shell-like structure induced by strong attraction. The stronger the attraction, the more particles can be packed on a specific shell, and the more shell-like the final structure will be. This is closely related to the phenomenon of surface tension where the attraction between molecules or atoms results in the minimization of the outer surface area. This competition results, in the case of strong attraction, in a sequential formation of regular polyhedra.

Our theoretical model may enable one to guide the synthesis of novel 3D assemblies in a controlled and efficient manner, which will be of importance for different scientific applications where specific 3D arrangements of nanoparticles are required, such as metamaterials or nanoparticle assemblies with optimized hot spot density.

**Acknoledgments**

This work was supported by the Flemish Science Foundation (FWO-Vl) and the Methusalem programme of the Flemish government. Computational resources were provided by HPC infrastructure of the University of Antwerp (CalcUA) a division of the Flemish Supercomputer Center (VSC). The authors acknowledge financial support from European Research Council ( ERC Advanced Grant # 24691-COUNTATOMS,  ERC Advanced Grant # 267867-PLASMAQUO, ERC Starting Grant # 335078-COLOURATOMS). The authors also appreciate financial support from the European Union under the Seventh Framework Program (Integrated Infrastructure Initiative N. 262348 European Soft Matter Infrastructure, ESMI).

SUPPORTING INFORMATION

# Self-organisation of highly symmetric nanoassemblies: a matter of competition

Jesus E. Galván-Moya[†,#], Thomas Altantzis[‡,#], Kwinten Nelissen[†], Francois M. Peeters[†], Marek Grzelczak[§,‖], Luis M. Liz-Marzán[§,‖], Sara Bals[‡,*], Gustaaf Van Tendeloo[‡]

[†]Department of Physics, University of Antwerp, Groenenborgerlaan 171, B-2020 Antwerp, Belgium

[‡]EMAT, University of Antwerp, Groenenborgerlaan 171, B-2020 Antwerp, Belgium

[§]Bionanoplasmonics Laboratory, CIC biomaGUNE, Paseo de Miramón 182, 20009 Donostia-San Sebastián, Spain

[‖]Ikerbasque, Basque Foundation for Science, 48011 Bilbao, Spain

## AUTHOR INFORMATION

### # Author contribution

J.E.G-M and T.A. equally contributed to this work.

### * Corresponding author

Email: sara.bals@uantwerpen.be

The authors declare no competing financial interest

## S1. Methodology section

Synthesis: Nine batches of gold nanoparticles (18.0 ± 0.5 nm, 40.0 ± 0.9 nm and 61.7 ± 1.5nm), stabilized with polystyrene (Mw = 5.8, 21.5, 53 kg/mol) were prepared according to experimental conditions reported in Reference 1. In a typical assembly experiment, water (0.4 mL) was added to the polystyrene-stabilized gold colloid (1.6 mL, THF) under magnetic stirring. After 10 minutes, a solution of polystyrene-block-polyacrylic acid was added in THF (6 mg/mL, 0.2 mL). Subsequently, the water content was increased up to 35 wt %, followed by increasing the temperature to 70 ºC, which was maintained for 1 h. The final solution was centrifuged twice and dispersed in pure water. As-prepared clusters were used for imaging without further processing.

Structural analysis: Electron microscopy observations were carried out using a FEI Tecnai G2 electron microscope operated at 200 kV. A Fischione tomography holder (model 2020) was used for the acquisition of the tilt series of 2D projection images. All tilt series were acquired in High Angle Annular Dark Field Scanning Transmission Electron Microscopy (HAADF-STEM) mode with an annular range from $-74^{o}$ to $+76^{o}$ and a tilt increment of $2^{o}$. The alignment of the series was performed in Inspect 3D software (FEI). All the reconstructions were performed using the Simultaneous Iterative Reconstruction Technique (SIRT), as implemented in Inspect 3D.

## S2. Structures of the highly symmetric nanoassemblies

Using 3D tomographic reconstruction we visualized the structural organization of gold nanoassemblies, as shown in Figure S1. Figures S1a-c show a few examples of different assemblies that were experimentally found, evidencing a spherically organized arrangement into shells for each of these assemblies. Figures S1d-f show only the particles located at the inner shells. In Figures S1d,e the arrangement of the inner shells is found to be a tetrahedron and an icosahedron, for systems of N=33 and N=59 particles, respectively. These reconstructions evidence the highly symmetric arrangement of particles in this kind of systems.

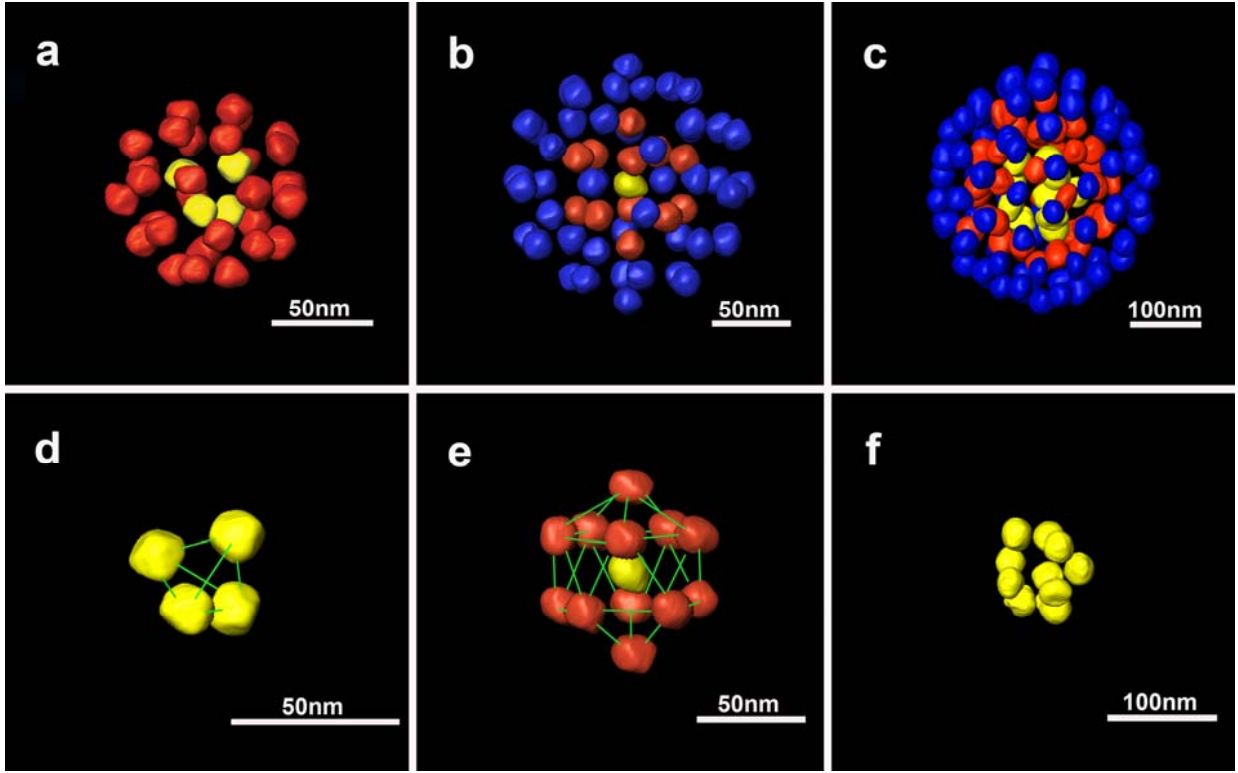

**Figure S1.** Tomographic reconstructions of the experimentally obtained assemblies. a, b, c) 3D representations of three different reconstructed assemblies. d, e, f) 3D representation of the inner shell of the assemblies presented in Figures S1 a-c respectively, showing their regular structures e.g. the tetrahedron (d) for N=33 and the icosahedron (e) for N=59.

## S3. Theoretical analysis

The interaction between particles is complex and involves many variables, as was discussed in Reference 1, where the described system consists of polystyrene (PS)-stabilized spherical gold nanoparticles (Au) dispersed in tetrahydrofuran (THF), and the interaction energy between each pair of particles was assumed as follows:

$$E = \frac{100DL^2kT}{\pi s^3}e^{-\pi r_D/L} - \frac{D}{12}\left(\frac{A_{232}}{r_D - 2L} - \frac{2A_{123}}{r_D - L} + \frac{A_{121}}{r_D}\right) - 2\pi DD_0\gamma(1-f)e^{-(r_D - 2L)/D_0}. \quad (S3.1)$$

In this expression $r_D$ is the distance between the surfaces of two interacting nanoparticles, D and L are the particle diameter and the polymer chain length respectively, while A is the Hamaker constant ($A_{123}=A_{Au-PS-THF}$, $A_{121}=A_{Au-PS-Au}$ and $A_{232}=A_{PS-THF-PS}$). The hydrophobic interaction is described by the dimensionless parameter $f$. The parameters listed in Table S1 are taken from Reference 1.

| Symbol | Description | Value |
|---|---|---|
| **D** | Diameter of nanoparticle | 20.0 nm |
| **L** | Average of the polymer chain length | 39.0 nm |
| **s** | Footprint of the polymer chain length | 1.4 nm |
| **T** | Temperature of the sample | 298 K |
| **$A_{123}$** | Hamaker constant | $2.3 \times 10^{-20}$ J |
| **$A_{121}$** | Hamaker constant | $1.0 \times 10^{-19}$ J |
| **$A_{232}$** | Hamaker constant | $5.7 \times 10^{-21}$ J |
| **$D_0$** | Decay length for the hydrophobic force | 1.0 nm |
| **$\gamma$** | Interfacial tension of the polyestyrene-20%water/80%THF | $8.0 \times 10^{-3}$ J/m$^2$ |

**Table S1.** List of the synthesis parameters used in the model of Equation (S3.1).

It was also described earlier that in the case of sufficiently large hydrophobic interactions, two energetically favorable stability regions are found for two interacting particles. These regions are highlighted in Figure S2. The first region is located around $r_1 = D + 2L$ whereas the second region can be found around the second minimum for $r_2 > r_1$. A similar landscape for such pairwise interaction was found in Reference 2, by considering competing interactions. These two stability regions (local minima) are separated by an energy barrier, which is modulated by the hydrophobic interaction. One could conclude that, in case the average inter-particle distance is found inside the first region, the system is characterized by a densely packed configuration, whereas if the average distance between particles is found inside the second region, the system will be arranged in a shell-like configuration. Previous models[1,2] show that the inter-particle distance of the assemblies, is around the second minimum ($r_2$).

From Reference 1 it is clear that the presence of the hydrophobic attraction decreases the height of the energetic barrier between these two regions, thereby increasing the probability to form densely packed configurations. On the other hand, when the effective contribution from the hydrophobic interaction is small, which is the case for long polymer chains compared to the particle size, the distances between particles are large enough to form shell-like structures.

The model used in Reference 1 allows a good description of the inter-particle distance, i.e. between two free particles. However, a physical understanding of the self-assembly process cannot be extracted from that model, because the systems found for different synthesis parameters cannot form large stable clusters. Therefore, we propose a potential that allows us to fit the landscape of Equation (S3.1) around the second local minimum, but which also allows us to tune the interaction range. A potential satisfying these conditions is given by:

$$E = \sum_{i=1}^{N} \sum_{j>i}^{N} \left( A \exp(-\alpha\, r_{ij}) - \tilde{B} \exp(-\tilde{\beta}\, r_{ij}) \right). \tag{S3.2}$$

In this expression A and $\tilde{B}$ modulate the strength while $\alpha$ and $\tilde{\beta}$ are a measure of the screening of the repulsion and attraction between two particles, respectively. $N$ corresponds to the number of particles of the assembly and $r_{ij}$ represents the distance between the center of the $i$-th and the center of the $j$-th particle in the self-assembled system. Please note that for $B = 2/Ae$, $\alpha = 2$ and $\beta = 1$ one gets the well known Morse Potential, which has been studied in detail.[3] Equation (S3.2) can be written in dimensionless form as follows:

$$E = \sum_{i=1}^{N} \sum_{j>i}^{N} \left( \exp(-r_{ij}) - B \exp(-\beta\, r_{ij}) \right), \tag{S3.3}$$

where $\beta = \tilde{\beta}/\alpha$, while the energy and the distances are given in units of $E_0 = AK_BT$ and $r_0 = \alpha \tilde{r}_0$, respectively. The average distance between shells in the assembly is given by $\tilde{r}_0$, which is the characteristic distance of the system. This potential consists of a repulsive and an attractive term, where the latter is modulated by its strength B, and interaction range $1/\beta$. A similar interaction has been used to describe successfully the structural formation of 2D self-organized colloidal systems.[4] Here, we expand the approach to 3D assemblies.

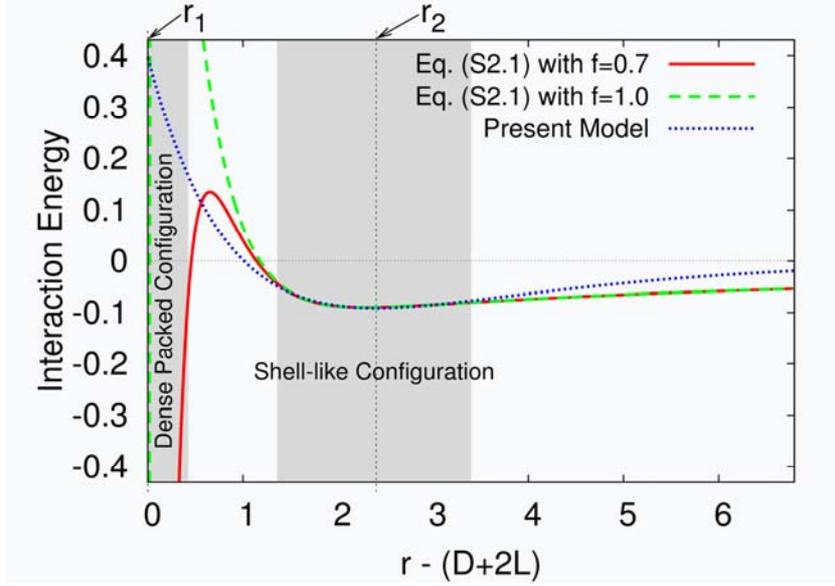

**Figure S2.** Interaction energy between two particles as a function of the separation distance. The red (solid) and green (dashed) curves represent the energy landscape of Equation (S3.1) for different hydrophobic factors (f), and the blue curve shows the interaction considered in the present model wtih $B = 0.62$ and $\beta = 0.5$. Energy and distance are expressed in dimensionless units.

The relation between synthesis parameters and our theoretical model is obtained by fitting Equation (S3.1) with Equation (S3.2) around the second minimum, using all synthesis parameters listed in Table S1. Based on this fit, the parameters of our model are given in Table S2, where β=0.5 was found to be an optimal screening parameter for all investigated samples, the average inter-shell distance ($r_0$) is also listed as it was extracted from experimental samples.

| Sample | $r_0$ (nm) | A | $\alpha$ | B |
|---|---|---|---|---|
| A | 27 ± 2 | 2.2150 | 0.0390 | 0.5982 |
| B | 33 ± 3 | 0.9541 | 0.0162 | 0.6183 |
| C | 53 ± 7 | 1.4920 | 0.0131 | 0.6433 |

**Table S2.** The average distance between shells ($r_0$) extracted from experiment, and the values of the theoretical parameters A, $\alpha$ and B are listed for different samples. The screening parameter considered in the fitting process is $\beta=0.5$.

The correction introduced by the present model is represented by the screening of the pairwise interaction potential after the minimum, as shown in Figure S2 for $r-(D+2L)>3$. Note that the structure of the configurations is controlled by only two independent parameters B and $\beta$ since $r_0$ and $\alpha$ determines the length scale of the sample, and A the energy scale.

In order to understand the cluster formation process, in Table S3 we present the cluster configuration for all metastable states as a function of the number of particles in the cluster for B=0.65 and $\beta=0.5$. In this table the energy per particle for each configuration is presented and the percentage difference with respect to the ground state is calculated. Notice that for small number of particles only one stable state is found, but the number of metastable states increases with the number of particles in the cluster. However, about the metastable states it is important to remark that the energy difference with respect to the ground state is less than 0.1% for N>20, which make it experimentally likely to find a metastable state rather than the ground state configuration.

| N | Configuration | | | E/N | %$E_0$ |
|---|---|---|---|---|---|
| 3 | 0 | 0 | 3 | -0.10563 | |
| 4 | 0 | 0 | 4 | -0.15844 | |
| 5 | 0 | 0 | 5 | -0.20740 | |
| 6 | 0 | 0 | 6 | -0.25798 | |
| 7 | 0 | 0 | 7 | -0.30428 | |
| 8 | 0 | 0 | 8 | -0.35148 | |
| 9 | 0 | 0 | 9 | -0.39785 | |
|   | 0 | 1 | 8 | -0.39447 | 0.85097 |
| 10 | 0 | 0 | 10 | -0.44339 | |
|    | 0 | 1 | 9 | -0.44183 | 0.35109 |
| 11 | 0 | 1 | 10 | -0.48828 | |
|    | 0 | 0 | 11 | -0.48770 | 0.1202 |
| 12 | 0 | 0 | 12 | -0.53383 | |
|    | 0 | 1 | 11 | -0.53349 | 0.064 |
| 13 | 0 | 1 | 12 | -0.58021 | |
|    | 0 | 0 | 13 | -0.57681 | 0.58615 |
| 14 | 0 | 1 | 13 | -0.62395 | |
|    | 0 | 0 | 14 | -0.62083 | 0.49945 |
| 15 | 0 | 1 | 14 | -0.66854 | |
|    | 0 | 0 | 15 | -0.66429 | 0.63605 |

| | | | | | | | | | | | |
|---|---|---|---|---|---|---|---|---|---|---|---|
| 16 | 0 | 1 | 15 | -0.71254 | | | 0 | 4 | 26 | -1.31833 | 0.00965 |
| | 0 | 2 | 14 | -0.71100 | 0.21497 | 31 | 0 | 6 | 25 | -1.36127 | |
| | 0 | 0 | 16 | -0.70766 | 0.68465 | | 0 | 5 | 26 | -1.36126 | 0.00085 |
| 17 | 0 | 1 | 16 | -0.75639 | | | 0 | 4 | 27 | -1.36112 | 0.01074 |
| | 0 | 2 | 15 | -0.75522 | 0.15367 | 32 | 0 | 6 | 26 | -1.40422 | |
| 18 | 0 | 1 | 17 | -0.79995 | | | 0 | 5 | 27 | -1.40418 | 0.00285 |
| | 0 | 2 | 16 | -0.79946 | 0.06135 | | 0 | 4 | 28 | -1.40367 | 0.0395 |
| 19 | 0 | 1 | 18 | -0.84334 | | 33 | 0 | 6 | 27 | -1.44723 | |
| | 0 | 2 | 17 | -0.84330 | 0.00366 | | 0 | 5 | 28 | -1.44681 | 0.0291 |
| | 0 | 3 | 16 | -0.84199 | 0.15997 | | 0 | 4 | 29 | -1.44593 | 0.08984 |
| 20 | 0 | 2 | 18 | -0.88701 | | 34 | 0 | 6 | 28 | -1.48993 | |
| | 0 | 1 | 19 | -0.88610 | 0.10264 | | 0 | 7 | 27 | -1.48955 | 0.02525 |
| 21 | 0 | 2 | 19 | -0.93021 | | | 0 | 5 | 29 | -1.48923 | 0.04702 |
| | 0 | 3 | 18 | -0.92991 | 0.03236 | | 0 | 4 | 30 | -1.48837 | 0.10441 |
| | 0 | 1 | 20 | -0.92931 | 0.09607 | 35 | 0 | 6 | 29 | -1.53250 | |
| 22 | 0 | 2 | 20 | -0.97370 | | | 0 | 7 | 28 | -1.53232 | 0.01156 |
| | 0 | 3 | 19 | -0.97332 | 0.03867 | | 0 | 8 | 27 | -1.53194 | 0.0364 |
| | 0 | 1 | 21 | -0.97206 | 0.1683 | | 0 | 5 | 30 | -1.53181 | 0.04463 |
| 23 | 0 | 3 | 20 | -1.01700 | | 36 | 0 | 6 | 30 | -1.57513 | |
| | 0 | 2 | 21 | -1.01676 | 0.02333 | | 0 | 7 | 29 | -1.57497 | 0.01048 |
| | 0 | 4 | 19 | -1.01629 | 0.07012 | | 0 | 8 | 28 | -1.57475 | 0.02397 |
| 24 | 0 | 3 | 21 | -1.06023 | | | 0 | 5 | 31 | -1.57421 | 0.05836 |
| | 0 | 4 | 20 | -1.06011 | 0.01133 | 37 | 0 | 7 | 30 | -1.61769 | |
| | 0 | 2 | 22 | -1.05976 | 0.04399 | | 0 | 6 | 31 | -1.61766 | 0.00208 |
| 25 | 0 | 3 | 22 | -1.10345 | | | 0 | 8 | 29 | -1.61745 | 0.01533 |
| | 0 | 4 | 21 | -1.10337 | 0.00751 | | 0 | 5 | 32 | -1.61677 | 0.05683 |
| | 0 | 2 | 23 | -1.10230 | 0.10454 | 38 | 0 | 6 | 32 | -1.66031 | |
| 26 | 0 | 4 | 22 | -1.14672 | | | 0 | 7 | 31 | -1.66028 | 0.00154 |
| | 0 | 3 | 23 | -1.14619 | 0.04671 | | 0 | 8 | 30 | -1.66023 | 0.00446 |
| | 0 | 5 | 21 | -1.14582 | 0.07903 | 39 | 0 | 7 | 32 | -1.70299 | |
| | 0 | 2 | 24 | -1.14519 | 0.13398 | | 0 | 8 | 31 | -1.70291 | 0.00452 |
| 27 | 0 | 4 | 23 | -1.18963 | | | 0 | 9 | 30 | -1.70260 | 0.02291 |
| | 0 | 3 | 24 | -1.18928 | 0.02922 | | 0 | 6 | 33 | -1.70237 | 0.03634 |
| | 0 | 5 | 22 | -1.18919 | 0.03687 | 40 | 0 | 8 | 32 | -1.74566 | |
| 28 | 0 | 4 | 24 | -1.23281 | | | 0 | 9 | 31 | -1.74529 | 0.02113 |
| | 0 | 5 | 23 | -1.23225 | 0.04572 | | 0 | 7 | 33 | -1.74519 | 0.02725 |
| | 0 | 3 | 25 | -1.23183 | 0.08011 | | 0 | 6 | 34 | -1.74467 | 0.05672 |
| | 0 | 6 | 22 | -1.23181 | 0.08179 | 41 | 0 | 9 | 32 | -1.78809 | |
| 29 | 0 | 4 | 25 | -1.27556 | | | 0 | 8 | 33 | -1.78796 | 0.00706 |
| | 0 | 5 | 24 | -1.27554 | 0.00182 | | 0 | 7 | 34 | -1.78755 | 0.02974 |
| | 0 | 6 | 23 | -1.27500 | 0.04414 | 42 | 0 | 9 | 33 | -1.83047 | |
| | 0 | 3 | 26 | -1.27443 | 0.08874 | | 0 | 8 | 34 | -1.83040 | 0.00337 |
| 30 | 0 | 6 | 24 | -1.31845 | | | 0 | 10 | 32 | -1.83033 | 0.00759 |
| | 0 | 5 | 25 | -1.31838 | 0.00545 | | 0 | 7 | 35 | -1.82984 | 0.03401 |

| | | | | | | | | | | |
|---|---|---|---|---|---|---|---|---|---|---|
| 43 | 0 | 9 | 34 | -1.87295 | | 50 | 0 | 12 | 38 | -2.16965 | |
| | 0 | 10 | 33 | -1.87276 | 0.01032 | | 0 | 11 | 39 | -2.16958 | 0.00319 |
| | 0 | 8 | 35 | -1.87276 | 0.01038 | | 1 | 12 | 37 | -2.16952 | 0.00602 |
| | 0 | 7 | 36 | -1.87212 | 0.04428 | | 0 | 10 | 40 | -2.16945 | 0.009 |
| 44 | 0 | 9 | 35 | -1.91535 | | | 1 | 11 | 38 | -2.16943 | 0.00995 |
| | 0 | 10 | 34 | -1.91528 | 0.00389 | | 1 | 10 | 39 | -2.16939 | 0.01188 |
| | 0 | 8 | 36 | -1.91507 | 0.01471 | | 11 | 1 | 37 | -2.16935 | 0.01361 |
| | 1 | 10 | 33 | -1.91474 | 0.03186 | | 1 | 13 | 36 | -2.16913 | 0.02383 |
| 45 | 0 | 9 | 36 | -1.95775 | | | 0 | 9 | 41 | -2.16896 | 0.03185 |
| | 0 | 10 | 35 | -1.95774 | 0.0003 | 51 | 0 | 12 | 39 | -2.21205 | |
| | 0 | 8 | 37 | -1.95737 | 0.01927 | | 1 | 12 | 38 | -2.21198 | 0.00326 |
| | 0 | 11 | 34 | -1.95736 | 0.01984 | | 0 | 11 | 40 | -2.21184 | 0.00954 |
| | 1 | 10 | 34 | -1.95726 | 0.02498 | | 1 | 11 | 39 | -2.21176 | 0.0133 |
| | 1 | 9 | 35 | -1.95726 | 0.02516 | | 0 | 10 | 41 | -2.21166 | 0.01785 |
| 46 | 0 | 10 | 36 | -2.00022 | | | 1 | 10 | 40 | -2.21163 | 0.019 |
| | 0 | 9 | 37 | -2.00011 | 0.00544 | | 1 | 13 | 37 | -2.21160 | 0.02036 |
| | 0 | 11 | 35 | -1.99989 | 0.01627 | 52 | 1 | 12 | 39 | -2.25439 | |
| | 1 | 10 | 35 | -1.99976 | 0.0229 | | 0 | 12 | 40 | -2.25438 | 0.00071 |
| | 1 | 9 | 36 | -1.99971 | 0.02534 | | 1 | 11 | 40 | -2.25412 | 0.0123 |
| | 0 | 12 | 34 | -1.99969 | 0.02651 | | 0 | 11 | 41 | -2.25408 | 0.01386 |
| | 0 | 8 | 38 | -1.99956 | 0.03297 | | 1 | 13 | 38 | -2.25407 | 0.01435 |
| 47 | 0 | 10 | 37 | -2.04260 | | | 0 | 13 | 39 | -2.25398 | 0.01851 |
| | 0 | 11 | 36 | -2.04239 | 0.01015 | | 1 | 10 | 41 | -2.25389 | 0.02208 |
| | 0 | 9 | 38 | -2.04239 | 0.01027 | | 0 | 10 | 42 | -2.25379 | 0.02653 |
| | 1 | 10 | 36 | -2.04225 | 0.01678 | 53 | 1 | 12 | 40 | -2.29673 | |
| | 0 | 12 | 35 | -2.04225 | 0.01686 | | 0 | 12 | 41 | -2.29663 | 0.00434 |
| | 1 | 9 | 37 | -2.04211 | 0.02384 | | 1 | 13 | 39 | -2.29648 | 0.01093 |
| | 1 | 11 | 35 | -2.04208 | 0.02542 | | 0 | 13 | 40 | -2.29641 | 0.01395 |
| 48 | 0 | 10 | 38 | -2.08494 | | | 1 | 11 | 41 | -2.29639 | 0.01474 |
| | 0 | 11 | 37 | -2.08481 | 0.00649 | | 0 | 11 | 42 | -2.29628 | 0.01955 |
| | 0 | 12 | 36 | -2.08476 | 0.00855 | | 1 | 10 | 42 | -2.29608 | 0.0282 |
| | 1 | 10 | 37 | -2.08467 | 0.01323 | | 0 | 10 | 43 | -2.29586 | 0.03771 |
| | 0 | 9 | 39 | -2.08463 | 0.01501 | 54 | 1 | 12 | 41 | -2.33900 | |
| | 1 | 11 | 36 | -2.08459 | 0.01695 | | 1 | 13 | 40 | -2.33891 | 0.00369 |
| | 1 | 12 | 35 | -2.08458 | 0.0171 | | 0 | 12 | 42 | -2.33887 | 0.00555 |
| | 1 | 9 | 38 | -2.08444 | 0.0238 | | 0 | 13 | 41 | -2.33868 | 0.01382 |
| 49 | 0 | 10 | 39 | -2.12724 | | | 1 | 14 | 39 | -2.33866 | 0.01458 |
| | 0 | 11 | 38 | -2.12721 | 0.00129 | | 1 | 11 | 42 | -2.33863 | 0.01568 |
| | 0 | 12 | 37 | -2.12720 | 0.00164 | | 0 | 11 | 43 | -2.33839 | 0.02586 |
| | 1 | 12 | 36 | -2.12709 | 0.00694 | | 0 | 10 | 44 | -2.33800 | 0.04292 |
| | 1 | 10 | 38 | -2.12706 | 0.00818 | 55 | 1 | 12 | 42 | -2.38127 | |
| | 1 | 11 | 37 | -2.12701 | 0.01075 | | 1 | 13 | 41 | -2.38119 | 0.00319 |
| | 0 | 9 | 40 | -2.12677 | 0.02201 | | 1 | 14 | 40 | -2.38108 | 0.00797 |
| | 1 | 9 | 39 | -2.12674 | 0.02337 | | 0 | 12 | 43 | -2.38105 | 0.00907 |

| N | | | | | N | | | | |
|---|---|---|---|---|---|---|---|---|---|
| | 0 | 13 | 42 | -2.38094 | 0.0135 | | 0 | 13 | 46 | -2.54962 | 0.02585 |
| | 0 | 14 | 41 | -2.38077 | 0.02083 | | 0 | 12 | 47 | -2.54935 | 0.03632 |
| | 1 | 11 | 43 | -2.38076 | 0.02134 | **60** | 1 | 15 | 44 | -2.59245 | |
| | 1 | 15 | 39 | -2.38069 | 0.02429 | | 1 | 14 | 45 | -2.59244 | 0.00026 |
| | 1 | 10 | 44 | -2.38036 | 0.03802 | | 1 | 16 | 43 | -2.59226 | 0.00732 |
| **56** | 1 | 12 | 43 | -2.42347 | | | 1 | 13 | 46 | -2.59222 | 0.00908 |
| | 1 | 13 | 42 | -2.42347 | 0.00023 | | 1 | 12 | 47 | -2.59190 | 0.0211 |
| | 1 | 14 | 41 | -2.42342 | 0.00239 | | 0 | 14 | 46 | -2.59190 | 0.02135 |
| | 0 | 12 | 44 | -2.42327 | 0.00847 | | 0 | 13 | 47 | -2.59167 | 0.03026 |
| | 0 | 13 | 43 | -2.42317 | 0.01262 | | 0 | 12 | 48 | -2.59148 | 0.03747 |
| | 1 | 15 | 40 | -2.42313 | 0.01419 | **61** | 1 | 15 | 45 | -2.63467 | |
| | 0 | 14 | 42 | -2.42307 | 0.01651 | | 1 | 14 | 46 | -2.63457 | 0.00353 |
| **57** | 1 | 12 | 44 | -2.46572 | | | 1 | 16 | 44 | -2.63457 | 0.00378 |
| | 1 | 13 | 43 | -2.46571 | 0.0003 | | 1 | 13 | 47 | -2.63429 | 0.01416 |
| | 1 | 14 | 42 | -2.46571 | 0.00048 | | 1 | 17 | 43 | -2.63423 | 0.01678 |
| | 1 | 15 | 41 | -2.46555 | 0.007 | | 1 | 12 | 48 | -2.63404 | 0.02375 |
| | 0 | 13 | 44 | -2.46543 | 0.0118 | | 0 | 14 | 47 | -2.63399 | 0.02573 |
| | 0 | 12 | 45 | -2.46534 | 0.01549 | | 0 | 13 | 48 | -2.63382 | 0.03213 |
| | 0 | 14 | 43 | -2.46533 | 0.01571 | **62** | 1 | 15 | 46 | -2.67681 | |
| **58** | 1 | 14 | 43 | -2.50799 | | | 1 | 16 | 45 | -2.67678 | 0.00123 |
| | 1 | 13 | 44 | -2.50798 | 0.00032 | | 1 | 14 | 47 | -2.67668 | 0.00483 |
| | 1 | 15 | 42 | -2.50786 | 0.00507 | | 1 | 17 | 44 | -2.67658 | 0.00885 |
| | 1 | 12 | 45 | -2.50781 | 0.00717 | | 1 | 13 | 48 | -2.67645 | 0.01358 |
| | 0 | 14 | 44 | -2.50763 | 0.0143 | | 0 | 14 | 48 | -2.67619 | 0.02341 |
| | 0 | 13 | 45 | -2.50754 | 0.01777 | **63** | 1 | 16 | 46 | -2.71902 | |
| | 1 | 16 | 41 | -2.50753 | 0.01844 | | 1 | 15 | 47 | -2.71895 | 0.00271 |
| | 0 | 12 | 46 | -2.50738 | 0.02421 | | 1 | 14 | 48 | -2.71888 | 0.00508 |
| **59** | 1 | 14 | 44 | -2.55028 | | | 1 | 17 | 45 | -2.71884 | 0.00653 |
| | 1 | 15 | 43 | -2.55018 | 0.00412 | | 1 | 13 | 49 | -2.71843 | 0.02166 |
| | 1 | 13 | 45 | -2.55011 | 0.00666 | | 0 | 15 | 48 | -2.71838 | 0.02353 |
| | 1 | 16 | 42 | -2.54991 | 0.01471 | | 0 | 14 | 49 | -2.71817 | 0.03126 |
| | 1 | 12 | 46 | -2.54989 | 0.01519 | | | | | | |
| | 0 | 14 | 45 | -2.54978 | 0.01949 | | | | | | |

**Table S3.** The Mendeleev-like table for the configuration of the assemblies found for B=0.65 and β=0.5 as function of the number of particles in the cluster (N). The lowest energy metastable states are presented, with their energy per particle (E/N) and percentage difference with respect to the ground state (%$E_0$) in the rightmost columns.

## S4. Material Dependence

The Hamaker constants modulate the van der Waals interactions in Equation (S3.1). These constants are related to the chemical compositions of the synthesis (solvent and surfactant).[1] In order to understand the effect of the synthesis composition on the attraction between particles, we show in Figure S3 the behavior of the parameter B, fitted from Equation (S3.1), as a function of the Hamaker constants $A_{123}$ (red squares), $A_{232}$ (blue triangles) and $A_{121}$ (green circles). In each case, when a Hamaker constant is varying, the rest of the parameters are fixed to the values presented in Table S1.

From Figure S3 one can see that the strength of the attraction (B) increases when the interaction between the nanoparticle and the solvent ($A_{123}$) increases. On the other hand, B decreases when $A_{232}$ increases, showing a strong dependence on the relation between the polymer and the solvent. A similar behavior is evidenced by increasing $A_{121}$, indicating that the attraction decreases when the Hamaker constant, related to the interaction between nanoparticles and polymer chains, increases.

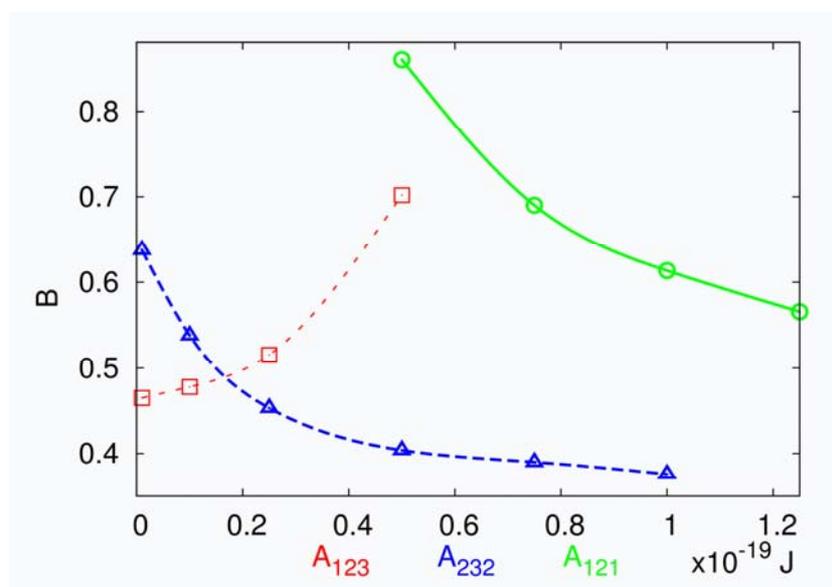

**Figure S3.** Theoretical parameter B plotted as a function of the Hamaker constants $A_{123}$ (red squares), $A_{232}$ (blue triangles) and $A_{121}$ (green circles). These points were calculated by fitting the interparticle energy Equation (S3.1), using the present model with a fix screening parameter of $\beta = 0.5$. Curves connecting symbols indicate

the behavior of a continuous increase of the Hamaker constant. All synthesis parameters used in this calculation are shown in Table S1.